# Magnetism driven ferroelectricity above liquid nitrogen temperature in $Y_2CoMnO_6$

G. Sharma, J. Saha and S. Patnaik[a]

School of Physical Sciences, Jawaharlal Nehru University, New Delhi-110067, India.

We report multiferroic behavior in double perovskite $Y_2CoMnO_6$ with ferroelectric transition temperature $T_c$ = 80K. The origin of ferroelectricity is associated with magnetic ordering of $Co^{2+}$ and $Mn^{4+}$ moments in a ↑↑-↓↓ arrangement. The saturation polarization and magnetization are estimated to be 65 $\mu C/m^2$ and 6.2 $\mu_B$/f.u. respectively. The magneto-electric coupling parameter, on the other hand, is small as a 5 Tesla field suppresses the electric polarization by only ~8%. This is corroborated with observed hysteretic behaviour at 5K that remains unsaturated even upto 7 Tesla. A model based on exchange-striction is proposed to explain the observed high temperature ferroelectricity.

[a]spatnaik@mail.jnu.ac.in

Magnetic control of electric polarization or electric control of magnetization has long been a challenging problem in condensed matter physics.[1-3] In the recent past the search for materials with such properties has received invigorated interest because of numerous potential applications that have been envisaged.[4-6] Moreover, with the advent of spin frustrated multiferroics, where-in ferroelectricity is driven by exotic magnetic order rather than non-centrosymmetic crystal structure, many novel material systems have come to the fore.[7-9] But two major problems remain; the magnitude of polarization remains exceedingly small and the onset of magnetic ordering leading to emergence of polarization remains a low temperature phenomena. To address these challenges, double perovskite oxides with general formula $A_2BB'O_6$ are under intense investigation.[9-14] For example, a range of magnetic ordering and the consequent tuning of magneto-electric properties have been reported in $Lu_2MnCoO_6$,[9] $Bi_2NiMnO_6$,[12] and $Ca_3CoMnO_6$.[13,14] The mechanism of multiferroicty in these perovskites is quite different compared to frustrated magnets such as $TbMnO_3$,[15] $Ni_3V_2O_8$,[16,17] $Bi_2Fe_4O_9$,[18] and hexagonal $YMnO_3$[19] where the spontaneous polarization is achieved by non-collinear ordering of adjacent magnetic moments. The mechanism of such *improper* ferroelctricity relates to spin-orbit coupling and is explained by the inverse Dzyaloshinskii-Moria (DM) interaction.[2] On the other hand, magnetostriction driven ferrolectricity, which can be induced by collinear magnetic ordering, is specifically applicable to double perovskite manganites.[1,9,20] Currently the double perovskites are viewed as the most promising candidates towards achieving large magnitude of induced polarization with ordering temperature above 40 K. Towards this end, in this letter, we report synthesis and characterization of an oxide double perovskite $Y_2CoMnO_6$ and confirm ferroelectric transition above the industrial benchmark of liquid nitrogen temperature.

Manganites with double perovskite structure are theoretically predicted to be multiferroic systems.[20] Here the ground state of magnetic structure can be tuned from ferromagnetic (La) to complex antiferromagnetic (Lu,Y) correlation with varying size of non-transition metal ion or rare earth ions.[20] Such magnetic orderings possess frustrated Ising spin chains with ↑↑-↓↓ spin pattern that are predicted to break the spatial inversion symmetry leading to the emergence of ferroelectricity.[14,20]

Polycrystalline samples of $Y_2CoMnO_6$ were synthesized by the solid state reaction method from the stoichiometric mixture of $Y_2O_3$, $CoO$ and $MnO_2$. It was ground and pelletized under pressure of 5 ton. The first sintering was done at 1100°C for 20 hour. It was then repeated at 1180°C after repeating the procedure.[21] The room temperature powder X-ray diffraction (XRD) pattern was collected in a PANalytical X'Pert PRO and Rietveld refinement was done with the GSAS software.[22] Pyroelectric current was measured with Keithley 6514 electrometer by warming the sample at a rate of 1K per min and the polarization was derived by integrating pyroelectric current over time. The magnetic property was studied using a *Quantum Design* MPMS and *Cryogenic* PPMS.

At high temperature, $Y_2CoMnO_6$ adopts a monoclinic crystal structure. As elucidated in Figure 1, the $Y^{3+}$ ions are located between two consecutive layers each of which is made up off tilted corner sharing octahedra around Co and Mn ions. In a sister compound $Lu_2CoMnO_6$, it is reported that $Co^{2+}$ and $Mn^{4+}$ ions form Co-Mn-Co-Mn chains along c-axis in an up-up-down-down (↑↑↓↓) ordering.[9] Similar orderings are also observed in $Y_2NiMnO_6$[20] and $Ca_3CoMnO_6$.[13,14] X-ray diffraction data and subsequent Reitveld refinement, as shown in Figure 2, confirm that the sample has been synthesized in single phase with monoclinic space group $P2_1/n$.[11] The room temperature cell parameters of $Y_2CoMnO_6$, as obtained from the GSAS software,[22] are given by **a** = 5.2322(2) Å, **b** = 5.5901(2) Å, **c** = 7.4685(3) Å, α = 90.00°, β = 89.92°(4) and γ = 90.00°. The atomic coordinates, bond lengths and bond angles are also estimated from the room temperature XRD and are summarized in Table I. The monoclinic crystal structure of $Y_2CoMnO_6$ contains two distinguishable positions for $Co^{2+}$ and $Mn^{4+}$ and three in-equivalent positions for oxygen atoms ($O_1$, $O_2$, $O_3$). The bond lengths are estimated to vary from 2.0404 to 2.0588 Å for Co-O and 1.9134 to 1.9895 Å for Mn-O, which are slightly larger than the corresponding case of $Lu_2CoMnO_6$. The overall picture that emerges is that while the magnetic structure and ordering would be close to that of $Lu_2CoMnO_6$,[9] the octahedral distortions would be smaller in the case of $Y_2CoMnO_6$ supporting much higher magnetic transition temperature.[10]

To ascertain the magnetic properties, DC magnetization measurements as a function of temperature and field were performed under field cooled (FC) and zero field cooled (ZFC) protocols. Figure 3 shows magnetic behaviour of $Y_2CoMnO_6$ under 0.01 and 0.005 T. ZFC magnetization measurement indicates a sharp down turn ~ 80 K and the FC and ZFC curves separate at this temperature. At the lowest temperature, the FC branch tends towards saturation ~ 0.5 $\mu_B$ under 0.005 T. We note that with higher fields, the temperature marking the separation of ZFC and FC graphs goes to lower temperatures and the saturation moment increases. The inverse of susceptibility ($\chi^{-1} = (T-\theta)/C$) as plotted in the inset a of Figure 3 yields Curie-Weiss constant C = 4.89 emu $Oe^{-1}mole^{-1}K^{-1}$ and $\theta$ = 83 K. The positive Curie-Weiss temperature indicates long range canted-spin correlation and the extrapolated effective moment of 6.2 $\mu_B$ is close to the theoretical value corresponding to one $Co^{2+}$ (S=3/2) and one $Mn^{4+}$ (S=3/2) per formula unit. Though not shown here, hysteresis plots with coercivity ~ 2 T (at 5K) were obtained that exhibited unsaturated moment of ~3.5 $\mu_B$ per formula unit upto 7 Tesla. In the lower inset of Figure 3 we show AC susceptibility measurement as a function of temperature at 7 Hz, 77 Hz and 777 Hz. A long range magnetic ordering is clearly demarcated around 80K below which frequency dependent characteristics is observed. Such behaviour has been assigned to slow dynamics of domain wall movement between ↑↑ and ↓↓ ferromagnetic domains.[9] The domain wall slides are effective means to control the electric polarization as a function of magnetic field involving magneto-striction.[1,9]

We next discuss the ferroelectric properties of $Y_2CoMnO_6$. For the measurement of electrical polarization P, the sample was polled from 150 K to 2 K under electric field (E = 2KV/cm). In Figure 4 we plot the electric polarization as a function of temperature and magnetic field. A robust polarization is confirmed below the magnetic transition ~80 K with saturation value of polarization 65 $\mu C/m^2$ at zero external magnetic field which is comparable to the values reported in $Ca_3CoMnO_6$ ($T_c$ ~14 K).[13,14] We also observe that in the presence of 5 T external field, the saturation polarization decreases to ~ 60 $\mu C/m^2$. Inset a of Figure 4 shows the actual pyroelectric current data in warming cycle after the specimen was polled. The anomaly in dielectric constant and tan $\delta$ across the magnetic phase transition were also confirmed (inset 4b). The electrical resistivity of the specimen plotted as a function of temperature (inset 4c) emphasizes highly resistive behaviour, particularly at low temperatures. It is to be noted that our sample is polycrystalline and the measured saturation polarization would be somewhat average of polarization across various crystal axes.

While a comprehensive understanding of the magnetically driven high temperature ferroelctricity in $Y_2CoMnO_6$ would demand a detailed in-field neutron scattering analysis, in the following we propose a qualitative model based on our data. In $Lu_2CoMnO_6$ whose crystal structure is identical to $Y_2CoMnO_6$, the emergence of ferroelectricity is assigned to ferromagnetic domain boundary between alternating Co and Mn layers along c-axis. From the magnetic structure point of view, $Y_2CoMnO_6$ is similar to $Lu_2MnCoO_6$,[9] but larger atomic radii of Y yields to stronger magnetic interaction. As in the case with $Ca_3CoMnO_6$, it is now established that such ↑↑-↓↓ arrangement of $Co^{2+}$ and $Mn^{4+}$ along c-axis can lead to symmetry lowering atomic displacements along c- axis with inequivalent Co-Mn bonds between alternate pairs. Such non-centrosymmetric structural distortions due to exchange striction is the microscopic cause of ferroelctricity in otherwise centro-symmetric monoclinic crystal of $Y_2CoMnO_6$. A simplistic magneto-striction model in our case would mean that the domain boundaries between adjacent Co - Mn layers (along c-axis) would sustain a polarization that would be opposite in direction to that induced by Mn - Co layers but due to inequivalent cancellation, there would be effective non-zero polarization along c-axis. With the application of external magnetic field, the structural distortions caused by spin –phonon coupling would then correlate ferroelectricity to altered magnetic orderings. Moreover, we observe that the saturation magnetization from hysteresis measurement is 58% of theoretical value and the effect of magnetic field is therefore rather subdued to alter the magnetic state at 5 K. Under such scenario, it also supports our data regarding small suppression of polarization as a function of magnetic field. The magneto-electric coupling parameter ($\alpha = \Delta P/\Delta H$) is estimated to be 1.19 $\mu C/m^2 T$ which is comparable to the value obtained in magneto-striction driven mutiferroic $NdCrTiO_5$.[23] We note that collinear magnetic magnetic ordering of double perovskites would effectively nullify the DM interaction $\boldsymbol{P} = \boldsymbol{e}_{ij} \times \boldsymbol{S}_i \times \boldsymbol{S}_j$, where $\boldsymbol{e}_{ij}$ is the unit vector connecting sites with spin moment $S_i$ and $S_j$. Thus the spin-phonon coupling based striction mechanism as evidenced in multiferroicity of $Dy(Gd)FeO_3$[24] and $Ca_3CoMnO_6$[13,14] seems most appropriate to explain the observed ferrelectricity in $Y_2CoMnO_6$.

In conclusion, we have established magnetic structure driven ferroelectricity in double perovskite $Y_2CoMnO_6$ that manifests such properties above the liquid nitrogen temperature. Both the ferroic orders emerge simultaneously ~80 K. The ↑↑-↓↓ spin arrangement under long range magnetic correlation are understood to be the driving force for the emergence of magnetostriction driven ferroelectricity.

We acknowledge DST, India for SQUID at IIT Delhi and AIRF, JNU for access to PPMS and XRD measurements. We thank C.S. Yadav and G.C. Tiwari for useful discussion. G.S., J.S. acknowledges UGC and CSIR, India respectively for financial support.

Table I. Structural parameters of $Y_2CoMnO_6$

| Atom | X | Y | Z |
|---|---|---|---|
| Y | 0.5179(6) | 0.5734(2) | 0.2496(13) |
| Co | 0.0000 | 0.5000 | 0.0000 |
| Mn | 0.5000 | 0.0000 | 0.0000 |
| O1 | 0.3841 | 0.9585 | 0.2411 |
| O2 | 0.1971 | 0.1957 | -0.0575 |
| O3 | 0.3228 | 0.6953 | -0.0593 |
| Bond | | | Length (Å) |
| Co-O1 | | | 2.0404(1) |
| Co-O2 | | | 2.0348(1) |
| Co-O3 | | | 2.0588(1) |
| Mn-O1 | | | 1.9134(8) |
| Mn-O2 | | | 1.9735(5) |
| Mn-O3 | | | 1.9895(6) |
| Bond | | | Angle (Degree) |
| Co-O1-Mn | | | 141.6(2) |
| Co-O2-Mn | | | 145.5(1) |
| Co-O3-Mn | | | 142.0(1) |

**Figure Captions:**

**Figure: 1.** (Color online) The crystal structure of $Y_2CoMnO_6$. $Co^{2+}$ (Blue) and $Mn^{+4}$ (light green) ions are surrounded by oxygen polyhedra. $Y^{3+}$ and $O^{2-}$ are represented as light cyan and red spheres respectively. The Co-Mn-Co-Mn chains along c-axis form ↑↑-↓↓ spin arrangement.

**Figure 2.** (Color online) Room temperature X- ray powder diffraction pattern of $Y_2CoMnO_6$ and the Rietveld refinement profile. Observed (o), calculated (—), background (—) difference (—) are shown along with Bragg position (|). Monoclinic crystal structure in space group $P2_1/n$ is confirmed.

**Figure 3.** (Color online) Magnetization versus temperature at 0.005T (*) and 0.01T (o) measured under FC and ZFC protocol. Inset (a) shows $\chi^{-1}$ versus temperature with Curie - Weiss fitting. Inset (b) shows ac susceptibility measurement at 7, 77 and 777 Hz. Frequency dependent behaviour is observed below transition temperature.

**Figure 4.** (Color online) Electrical polarization is plotted as function of temperature in the presence of 0 and 5T external field. Inset a shows measured pyroelectric current as a function of temperature. Inset b marks observed anomaly tan (δ) across the magnetic phase transition. Inset c shows resistivity as a function of temperature implying excellent dielectric behavior.

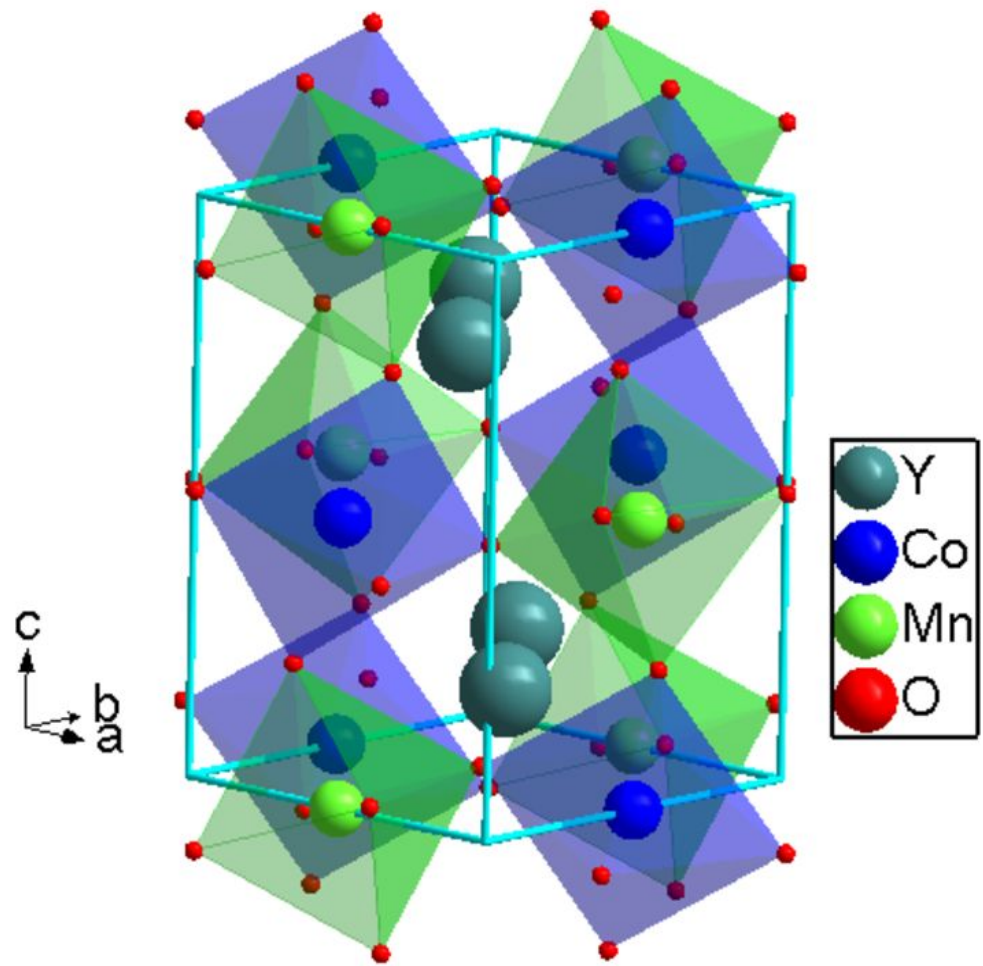

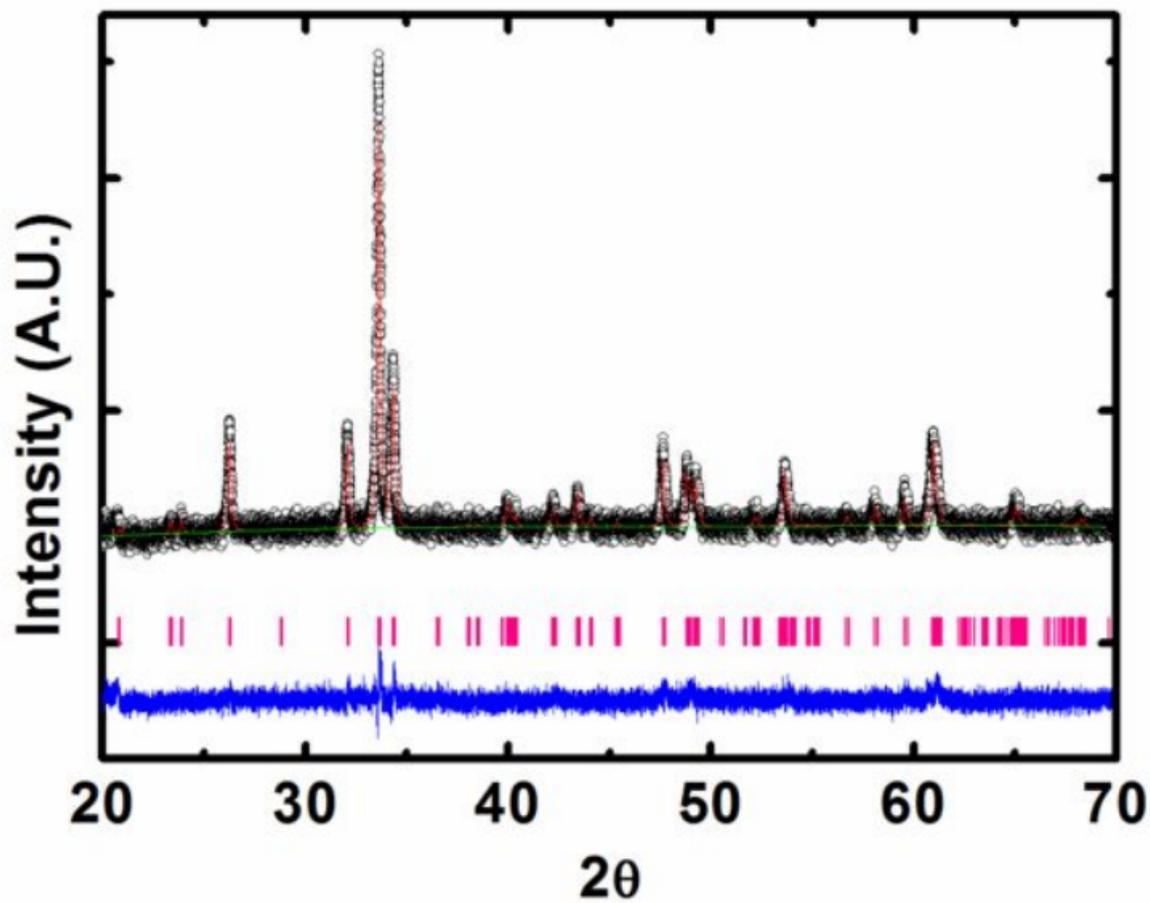

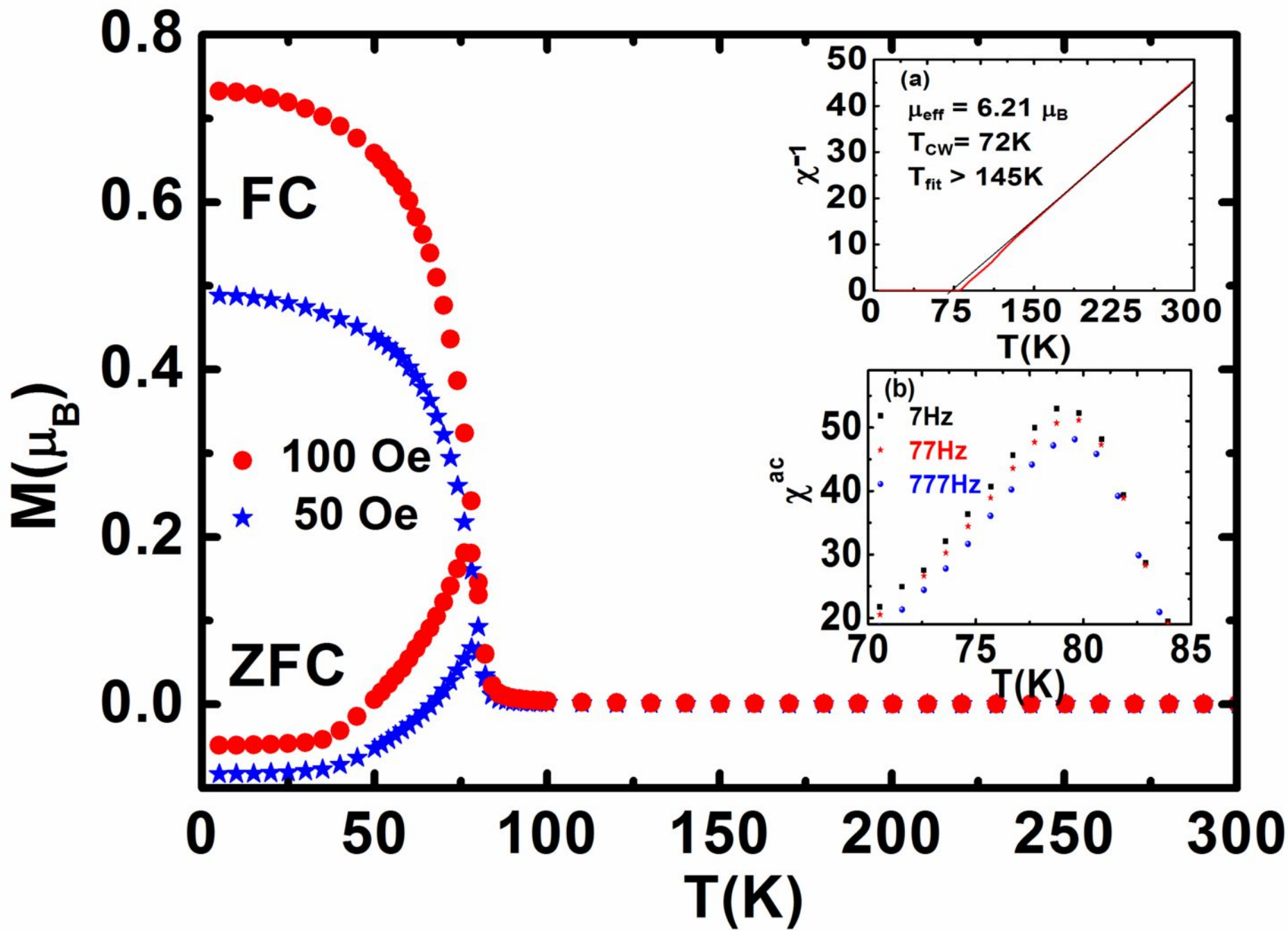

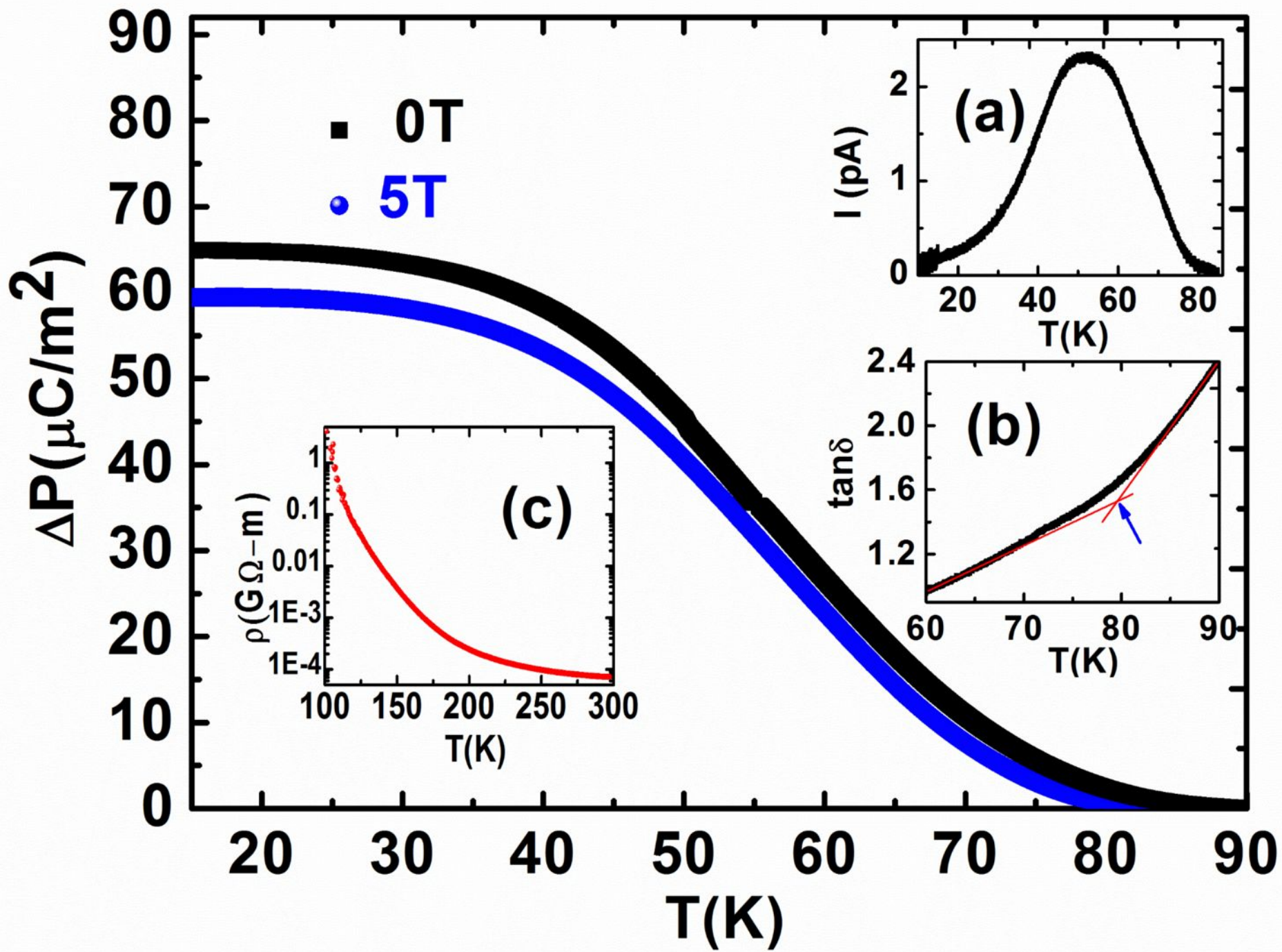